\documentclass[aps,twocolumn,showpacs]{revtex4-1}

\usepackage{graphicx}
\usepackage{subfigure}
\usepackage{amssymb}
\usepackage{amsfonts}
\usepackage{amsmath}
\usepackage{amsbsy}

\renewcommand{\vec}[1]{\mbox{\boldmath$\mathrm{#1}$}}

\begin{document}

\title{Piezoelectric control of the magnetic anisotropy via interface strain coupling in a composite multiferroic structure}

\author{Chenglong Jia$^{1,2}$, Alexander Sukhov$^{2,3}$, Paul P. Horley$^3$, and Jamal Berakdar$^2$}
\address{$^1$Key Laboratory for Magnetism and Magnetic Materials of the Ministry of Education, Lanzhou University, Lanzhou 730000, China\\
$^2$Institut f\"ur Physik, Martin-Luther Universit\"at Halle-Wittenberg, 06099 Halle (Saale), Germany\\
$^3$Centro de Investigaci\'{o}n en Materiales Avanzados (CIMAV S.C.), Chihuahua/Monterrey, 31109 Chihuahua, Mexico}

\begin{abstract}
We investigate  theoretically the magnetic dynamics in a ferroelectric/ferromagnetic heterostructure coupled via strain-mediated magnetoelectric interaction.  We predict an electric field-induced  magnetic switching in the
plane perpendicular to the  magneto-crystalline easy axis, and trace this effect back to
the piezoelectric control of the magnetoelastic coupling.  We also investigate  the magnetic remanence and the electric coercivity.
\end{abstract}


\maketitle

Controlling  magnetism with an electric field is
 fundamentally important and bears the potential for  a wide range of applications
as for instance in  sensorics  and  magnetoelectrically  controlled  spintronics devices with ultra low heat dissipation \cite{ME-memory, MTJ1,MTJ2}.
Multiferroics, \emph{i.e.}, materials that exhibit ferroic (magnetic, electric, and elastic) orders \cite{ME-review,Single-ME,Composite-ME},  allow to tune the coupled  ferroelectric (FE) and the ferromagnetic (FM) order parameters by external magnetic and electric fields, respectively. A major obstacle however is the
relatively low strength of the magnetoelectric coupling in bulk matter. A promising route to circumvent this problem
are the appropriately synthesized composite FE/FM nano and multilayer structures  that may serve as elements in
 quantitatively new multiferroic devices at room temperature \cite{Composite-ME, Zavaliche-11, MeKl11}.   One example that has been studied recently theoretically and experimentally is
  BaTiO$_3$(BTO)/Fe \cite{Tsymbal-06, Binek-07,Tsymbal-08,Taniyama-09,Berakdar-10,Bibes-11,MeKl11,Taniyama-12}.  Steering the magnetization of Fe layer in BTO/Fe by an electric field was  demonstrated experimentally  in Refs. \cite{Taniyama-09,Bibes-11,MeKl11}.  Several mechanisms may underlay the magnetoelectric effect in composite multiferroic junctions, for example charge rearrangements  \cite{Tsymbal-06,Mertig-10,Demkov-10,MeKl11,mAnst}, strain effects  \cite{Binek-07,Taniyama-09,Taniyama-11,Bibes-11}, and exchange-bias \cite{Binek-05,Laukhin-06,Wu-10} have been
   studied as the key ingredients  for the  magnetoelectric coupling. In a previous study we investigated theoretically the dynamic response to an applied external field in a multiferroic chain with a linear magnetoelectric coupling that results from the electrostatic screening at the FM/FE interface \cite{Berakdar-10,Berakdar-11}. An important finding is that  for the material parameters corresponding to BTO/Fe  the total electric polarization and the net magnetization are  controllable by external magnetic and electric fields, respectively \cite{Berakdar-12}.

\begin{figure}[b]
\centering
\includegraphics[width=0.45 \textwidth]{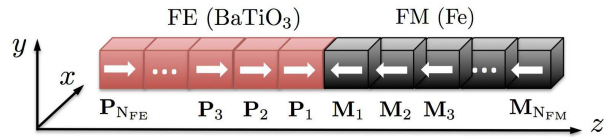}
\caption{Schematic view of the ferroelectric/ferromagnetic heterostructure.  The BTO film is assumed [001] oriented with the polarization perpendicular to the interface (pointing either $\hat{e}_z$ or $-\hat{e}_z$). The magnetic moments are taken with all three cartesian components and the initial directions are chosen randomly. }
\label{fig:chain}
\end{figure}

An outstanding important problem however is the question of  how the strain, due to the interface lattice distortion in a  FE/FM heterostructure, affects the multiferroic dynamics. Experimentally the thin FM Fe film on BTO single-crystal was successfully realized, and the magnetic properties of Fe film are found to be strongly modified by the successive structural transitions of BTO substrate via interfacial magnetoelastic coupling in three different phases, viz., tetragonal ($300$ K), orthorhombic ($230$ K), and rhombohedral ($150$ K) BTO phases  \cite{Binek-07,Taniyama-09,Taniyama-11,Taniyama-12}.  All these experimental finding are very promising steps towards the manipulating magnetic anisotropy using interfacial strain.  However,  the ferroelectric control of magnetism by the structure phase transition via the thermal cycle lacks high-speed read/write multiferroic dynamics as to their intrinsic relative long relaxation time. In contrast, the thin BTO film possesses strong piezoelectricity in experiments\cite{Bibes-11, Bruchhausen-08}. Epitaxial strain are found to enhance the spontaneous polarizations of the FE thin film.   Furthermore, as shown by Ederer and Spaldin in Ref.\cite{Piezoe1}, the strain dependence of the polarization in thin FE film can be understood in terms of the piezoelectric and the elastic constants of the unstrained materials.  In this letter,  we present  theoretical results for the interface strain-based magnetic response to an oscillating external electric field based on the piezoelectricity, instead of the thermal phase transition. The system under consideration is illustrated in Fig.\ref{fig:chain}.  An epitaxially grown multiferroic heterostructure  that we treat in a coarse-grained manner as consisting  of $N_{\text FM}$ ferromagnetic (e.g. Fe) and $N_{\text FE}$ ferroelectric (e.g. BaTiO$_3$) subcells. Each cell is a cube of equal volume $a^3 = 5 \times 5 \times 5 ~ \text{nm}^3$, as in Ref.\cite{Berakdar-12}.

%%%%%%%%%%%%%%%%%%%%%%%%%%%%%%%%%%%%%%%%%%%%%%%%%%%%%%%%%%
%%%%%%%%%%%%%%%%%%%%%%%%%%%%%%%%%%%%%%%%%%%%%%%%%%%%%%%%%%

For the epitaxially strained FE films,  the strain components  satisfy thus $u_4 = u_5 = u_6 =0$, $u_1 = u_2 \neq 0$,  and the strain perpendicular
to the interface is determined by the Poisson ratio $n = -u_1/u_3$, where $i=1,...,6$ denotes the strain tensor in Voigt notation.  Up to a linear order, the change in the polarization $\vec P$ is given by the improper piezoelectric tensor, $c_{\alpha i}=\frac{\partial P_{\alpha}}{\partial u_i}$ \cite{Piezoe2}. On the other hand, the magnetoelastic coupling at the interface is associated with an additional uniaxial anisotropy energy for the FM layer \cite{Book-FM1},
\begin{equation}
F_{MST} = - {3 \over 2} \lambda \sigma \cos^2 \phi
\end{equation}
where $\lambda$ is the average magnetostriction coefficient, and $\phi$ is the angle between the magnetization $\vec{M}$ and the direction of the stress $\vec{\sigma}$.  With these consideration of the
 stress-strain effect, the total free energy density of the multiferroic chain is
\begin{equation}
F_{MF} = F_{FE} + F_{FM}.
\end{equation}
The energy density of the FE subsystem is given by
\begin{equation}
F_{FE} = F_{G} + F_{DDI}
\end{equation}
where the elastic Gibbs function $F_{G}$ reads
\begin{eqnarray}
F_{G} = &-& \alpha \sum_{i} \vec P_i^2 + \beta \sum_{i}  \vec P_i^4 + \kappa \sum_{i} (\vec P_i -\vec P_{i-1})^2 \nonumber \\ &+&  c_{\text{eff}} u_1 \sum_i P_{iz} + {N_{FE} \over 2} C_u u_1^2 - \sum_{i} \vec{E}(t) \cdot \vec{P}_i.
\end{eqnarray}
Here $c_{\text{eff}} = 2c_{31} - c_{33}/n$ (note that the symmetry $c_{31} = c_{32}$ has been exploited to derived the effective piezoelectric constant \cite{Piezoe1}), $C_u$ is the stiffness coefficient of the FE film, and $\vec{E}(t) = (0,0, E_0 \sin \omega t)$ is an applied harmonic electric field. The long range FE dipole-dipole interaction $F_{DDI}$ has the usual form
\begin{equation}
F_{DDI} = \frac{1}{4\pi \epsilon_{\text {FE}} \epsilon_0}\sum_{i\neq k} \left[ \frac{\vec{P}_i \cdot \vec{P}_k - 3(\vec{P}_i \cdot \vec{e}_{ik})(\vec{e}_{ik}\cdot \vec{P}_k)}{r_{ik}^3} \right].
\end{equation}
where $\epsilon_{FE}$ is the FE permittivity,  $\epsilon_{0}$ is the dielectric constant of vacuum, $r_{ik}$ is the distance between  $\vec{P}_i$ and $\vec{P}_k$, and $\vec{e}_{ij}$ is the unit vector joining the two dipoles.

For  the ferromagnetic energy density the relation  applies
\begin{equation}
F_{FM} = F_{XC} + F_{MST} + F_{MMI}.
\end{equation}
$F_{XC}$ consists of the nearest-neighbor exchange interaction and the uniaxial magneto-crystalline anisotropy contributions
\begin{equation}
F_{XC} = - \frac{A}{a^2M_s^2} \sum_{j} \vec{M}_i \cdot \vec{M}_{j+1} - \frac{K_1}{M_s^2} \sum_{j} \vec{M}_{zj}^2
\end{equation}
with the saturation magnetization $M_s$. The magnetic dipole-dipole interaction is
\begin{equation}
F_{MMI} = \frac{\mu_0}{4\pi}\sum_{j\neq l} \left[ \frac{\vec{M}_j \cdot \vec{M}_l - 3(\vec{M}_j \cdot \vec{e}_{jl})(\vec{e}_{jl}\cdot \vec{M}_l)}{r_{jl}^3} \right]
\end{equation}
where $\mu_0$ is the magnetic permeability constant.

For the FE subsystem, the material parameters are chosen as $\alpha = 2.77 \times 10^7$ Vm/C \cite{Parameters-FE}, $\beta = 1.70 \times 10^8$ Vm$^5$/C$^3$ \cite{Parameters-FE}, $\kappa = 1.0 \times 10^8$ Vm/C \cite{Berakdar-10}, and $P_s = 0.499$ C/m$^2$ \cite{Berakdar-12}. The improper piezoelectric constants are set as those of BTO \cite{Piezoe1}: $c_{31} = 0.3$ $C/m^2$, $c_{33} = 6.7$ C/m$^2$, and the Poisson ratio $n=0.64$. The average stiffness coefficient $C_u$ of thin FE film is unknown and expected to quite smaller than that in the bulk \cite{mAnst}.  In the numerical simulation we used $C_u = 3.90 \times 10^9$ N/m$^2$ that is adjusted to reproduce the lattice mismatch ($\sim 1.4\%$) between the bcc Fe[001] film and the tetragonal BTO at room temperature \cite{Taniyama-12}.  Further material parameters concerning the  FM layer are  iron, i.e., $\lambda = 2.07 \times 10^{-5}$ along Fe [100] \cite{Book-FM1}, $A = 2.1 \times 10^{-11}$ J/m \cite{Book-FM2}, $K_1 = 4.8 \times 10^4$ J/m$^3$ \cite{Book-FM2}, and $M_s = 1.71 \times 10^6$ A/m \cite{Book-FM2}. The induced stress acting on the FM body is given by $\vec{\sigma} = - C_{11} (u_1, u_2, u_3)$ with $C_{11} = 1.78 \times 10^{11}$ N/m$^2$ being the elastic constant of BTO at the interface \cite{Book-FE}.

The multiferroic dynamics is studied by  Monte Carlo simulations with the standard Metropolis algorithm \cite{Book-MC} and open boundary condition at $T = 312$ K for tetragonal BTO phase.  The FE dipoles $\vec{P}_i$ are chosen to be uniformly distributed over $( 0, 0, -P_s)$-$(0,0,P_s)$ by a random function \cite{FE-MC1,FE-MC2}.  The three-dimensional magnetic moments $\vec{M}_i$ are updated coherently, \emph{i.e.}, at each trial Monte Carlo step a direction of new $\vec{M}^{\prime}_i$ is limited within a cone around the initial spin direction \cite{Hinzke-98}.  The maximum angle $\theta_{max}$ of the cone is determined by means of a feedback algorithm so that the number of accepted spin modifications is just half the total number of equilibrium configurations at temperature T and at time $t=0$ in the absence of the external electric field $\vec{E}(t)$  \cite{SGL-93}. After the multiferroic equilibrium at $T=312$ K is established,  $\theta_{max}$ is fixed and a sinusoidal oscillating electric field along the chain, $\vec{E}_z (t) = E_0 \sin  \omega t$, is turned on and sweep through the multiferrroic chain \cite{FM-MC1,FM-MC2,FM-MC3}. One sweep constitutes one Monte Carlo step (MCS),  which is taken as the time unit.  To reduce random errors, the data are collected and averaged for 400 full electric field cycles.

%%%%%%%%%%%%%%%%%%%%%%%%%%%%%%%%%%%%%%%%%%%%%%%%%%%%%%%%%%
\begin{figure}[b]
\centering
\includegraphics[width=0.48 \textwidth]{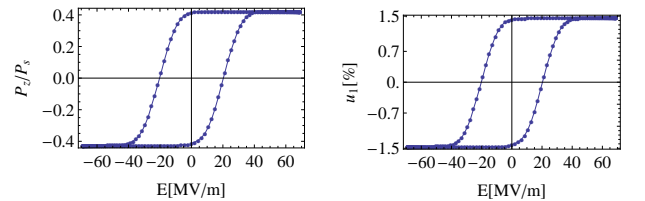}
\caption{FE polarization switching (left) and piezoresponse loop (right) driven by the harmonics electric field with period $2\pi/\omega = 200$ MCS. Hysteresis curves are calculated for $N_{FE} =20$ and $N_{FM} =4$.}
\label{fig:u-E}
\end{figure}
%%%%%%%%%%%%%%%%%%%%%%%%%%%%%%%%%%%%%%%%%%%%%%%%%%%%%%%%%%%%%%%%%%%%%%%%%%%%%%%%%%%%%%%%%%%%%%%%%%%%%%%%%%%%%%%%%%%%
\begin{figure}[t]
\centering
\includegraphics[width=0.48 \textwidth]{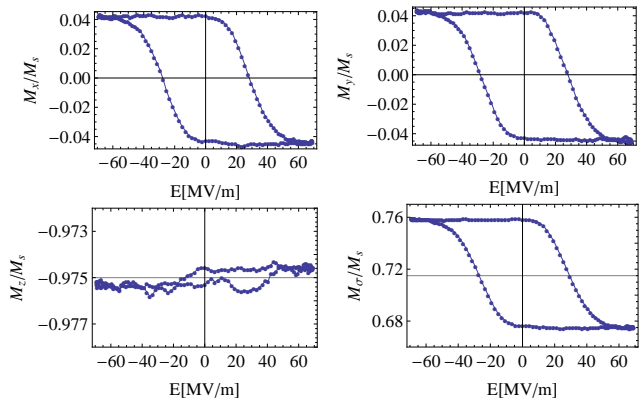}
\caption{Multiferroic response of the magnetic anisotropy to applied electric field. $M_{\sigma}$ is the projection component of magnetization in the direction of interface stress. The parameters are the same as in Fig.\ref{fig:u-E}.  }
\label{fig:M-E}
\end{figure}
%%%%%%%%%%%%%%%%%%%%%%%%%%%%%%%%%%%%%%%%%%%%%%%%%%%%%%%%%%
%%%%%%%%%%%%%%%%%%%%%%%%%%%%%%%%%%%%%%%%%%%%%%%%%%%%%%%%%%
\begin{figure}[b]
\centering
\includegraphics[width=0.48 \textwidth]{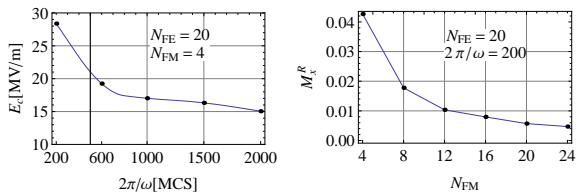}
\caption{The electric coercivity (left) and magnetic remanence (right) for $M_x$ hysteresis loop of the FM/FE heterostructure with $20$ FE cells. }
\label{fig:CR}
\end{figure}
%%%%%%%%%%%%%%%%%%%%%%%%%%%%%%%%%%%%%%%%%%%%%%%%%%%%%%%%%%

Fig.\ref{fig:u-E} shows the hysteretic piezoresponse of the FE subsystem. The FE layer is poled and expands or shrinks in response to the electric field.  Experimentally a similar piezoresponse has indeed been observed  in BTO/LSMO sample \cite{Bibes-11}.  In BTO film, the dynamics of strain is directly related to phonon modes of the crystal lattice. Their launching time in each cell is around $\sim 1$ ps, estimated with the sound velocity $v_{BTO} \approx 5000$ m/s at the room temerpature \cite{v-BTO},  which is much shorter compared to the oscillation period of the applied electric field \cite{Berakdar-10,Berakdar-11}.  We have thus a linear piezoelectric dynamic coupling of ferroelectric  and strain modes and time-independent Poisson ratio.  This piezoresponse in turn updates the induced interfacial stress, $\vec{\sigma} (t) = -C_{11} u_{1}(t) (1,1,-1/n)$, resulting in a time-oscillating uniaxial magnetic anisotropy, consequently an anisotropic multiferroic behavior in the FM film (c.f. Fig.\ref{fig:M-E}).  The maximum of  the stress anisotropy energy is estimated as $F_{MST} = 1.7 \times 10^{5}$ J/m$^3$, which gives rise to a pronounced magnetic response to the electric field in the normal-plane and along the direction of the stress $\vec{\sigma}$ even though the modification of the total magnetization is quite small.  As $u_1 > 0$,  we have $u_3 < 0$ that implies a tensile strain to Fe along the chain. The stress axis becomes magnetically harder, which favors a small alignment of the magnetization along the stress.  Whereas, $u_1 < 0$ indicates a compressive strain ($\lambda \sigma > 0$), that gives rise to an easy $\sigma$-axis and a large projection $M_{\sigma}$ (c.f. Fig.\ref{fig:M-E}.  Such an anisotropic magnetization behavior is qualitatively consistent with the experiment \cite{Binek-07,Taniyama-09,Taniyama-11}.
The change of in-plane magnetization reads $\Delta M_{\|}/M_{s} \approx 11\%$, smaller than that ($\sim 33 \%$) induced by the structure phase transition \cite{Binek-07}.  However, an advantage of the piezoelectric control is that the magnetic anisotropy can be changed by a high-frequency oscillating electric field instead of the slow thermal cycle.

It should be noted that the influence of the stress-strain effect on the magnetic anisotropy is typically more considerable than the surface magneto-crystalline anisotropy stemming from the charge redistribution at the FM/FE interface \cite{Tsymbal-06,Tsymbal-08,Mertig-10,Berakdar-10, mAnst}.  And the anisotropy introduced by the interfacial spin-polarized screening charges is along the chain \cite{Berakdar-10,Berakdar-11}, which results in a periodic variation of $M_x$ and $M_y$ but does not allow  to form the hysteresis loop \cite{Berakdar-12}.  To further characterize the electrical switchability of the  in-plane magnetization of the FM film, we study the frequency dependence and the finite-size effect on the magnetoelectric response, as shown Fig.\ref{fig:CR}.  The electric coercive fields for $M_x$ hysteresis loop are plotted as a function of the frequency of the applied electric field. Clearly, the increase in the frequency results in a decrease in the coercivity field $E_c$.  For thicker FM layer, the interface strain effect becomes weaker and is eventually  overwhelmed by the magneto-crystalline anisotropy, as marked by the fast decrease  of the remanent magnetization $M_{x}^{R}$  with $N_{FM}$.

In conclusion, we demonstrated theoretically that  the interface strain effect can strongly influence the multiferroic dynamics in FE/FM heterostructures. Through the piezoelectricity and the magnetostriction, the induced, electrically tunable magnetic anisotropy results in a well-developed  in-plane magnetic hysteresis loop as a function of the electric field.  This finding has a promising potential for the electric control of the complete magnetic switching whilst the total magnetization remains  stable in presence of applied voltage.  In order to get integrally insight into the multiferroic dynamics, the magnetoelectric coupling induced by the  screening charges should be taken into account as well.  For the strength of the magnetoelectric coupling in the range above $1$ s/F \cite{Berakdar-12},  two types of multiferroic interactions are comparable.  The stress anisotropy energy prefers the in-plane magnetization, whereas the surface spin-dependent screening charges favors to antiparallelly align the FE polarization and FM magnetization along the chain.   Studies of the competition between them promises rich potential for future multiferroic devices and are currently under way.

\acknowledgments{This work was supported by the Fundamental Research Funds for the Central Universities (Grant No. 860526, China), the National Natural Science Foundation of China (Grant No. 11104123), the National Basic Research Program of China (Grand. No. 2012CB933101),  the German Research Foundation (Grants No. SU 690/1-1 and No. SFB 762), and CONACYT, Mexico (Basic Science Projects No. 129269 and No. 133252).}


\begin{thebibliography}{30}
%
\bibitem{ME-memory} M. Bibes and A. Barth\'{e}l\'{e}my,  Nature Mater. \textbf{7}, 425 (2008).
%
\bibitem{MTJ1} M. Gajek,  M. Bibe, S. Fusil, K. Bouzehouane,  J. Fontcuberta,  A. Barth\'{e}l\'{e}my, and A. Fert, Nature Mater. \textbf{6}, 296 (2007).
%
\bibitem{MTJ2} D. Pantel, S. Goetze, D. Hesse and M. Alexe, Nature Mater. \textbf{DOI: 10.1038/NMAT3254}, (2012).
%
\bibitem{ME-review} W. Eerenstein, N. D. Mathur, and J. F. Scott,  Nature \textbf{442}, 759 (2006).
%
\bibitem{Single-ME} Y. Tokura and S. Seki,  Adv. Mater.  \textbf{22}, 1554 {2010}.
%
\bibitem{Composite-ME} Carlos A. F. Vaz, Jason Hoffman, Charles H. Ahn, and Ramamoorthy Ramesh,  Adv. Meter. \textbf{22}, 2900 (2010).
%
\bibitem{Zavaliche-11} F. Zavaliche, T. Zhao, H. Zheng, F. Straub, M. P. Cruz, P.-L. Yang, D. Hao, and R. Ramesh, Nano. Lett. \textbf{7}, 1586 (2007).
%
\bibitem{MeKl11} H.L. Meyerheim, F. Klimenta, A. Ernst, K. Mohseni, S. Ostanin, M. Fechner, S. Parihar, I.V. Maznichenko, I. Mertig, and J. Kirschner, Phys. Rev. Lett. {\bf 106}, 087203 (2011).
%
\bibitem{Tsymbal-06} C.-G. Duan, S. S. Jaswal, and E. Y. Tsymbal, Phys. Rev. Lett. \textbf{97}, 047201 (2006).
%
\bibitem{Binek-07} S. Sahoo, S. Polisetty, C.-G. Duan, Sitaram S. Jaswal, Evgeny Y.  Tsymbal, and C. Binek, Phys. Rev. B \textbf{76}, 092108 (2007).
%
\bibitem{Tsymbal-08} C.-G. Duan, J. P. Velev, R. F.  Sabirianov,  W. N. Mei,  S. S. Jaswal, and E. Y. Tsymbal,  Appl. Phys. Lett. \textbf{92}, 122905 (2008).
%
\bibitem{Berakdar-10} A. Sukhov, C.-L. Jia, P. P. Horley, and J. Berakdar, J. Phys. Condens.
Matter \textbf{22}, 352201 (2010).
%
\bibitem{Taniyama-09} T.  Taniyama, K. Akasaka, D.-S. Fu, and M. Itoh, J. Appl. Phys. \textbf{105}, 07D901 (2009).
%
\bibitem{Bibes-11} S. Valencia, A. Crassous, L. Bocher, V. Garcia, X. Moya,
R. O. Cherifi, C. Deranlot, K. Bouzehouane, S. Fusil,  A. Zobelli, A. Gloter, N. D. Mathur,  A. Gaupp,  R. Abrudan, F. Radu,  A. Barth\'{e}l\'{e}my, and M. Bibes, Nature Materials \textbf{10}, 753 (2011).
%
\bibitem{Taniyama-12} G. Venkataiah, Y. Shirahata, I. Suzuki, M. Itoh, and T. Taniyama, J.  Appl. Phys. \textbf{111}, 033921 (2012).
%
\bibitem{Mertig-10} M. Fechner, S. Ostanin, and I. Mertig,  J. Phys.: Conf. Ser.  \textbf{200}, 072027 (2010).
%
\bibitem{Demkov-10} J. Lee, Na Sai, T.Y. Cai, Q. Niu, and Alexander A. Demkov, Phys. Rev. B \textbf{81}, 144425 (2010)
%
\bibitem{mAnst} A. Mardana, Stephen Ducharme, and S. Adenwalla, Nano Lett. \textbf{11}, 3862 (2011).
%
\bibitem{Taniyama-11} G. Venkataiah, Y. Shirahata, M. Itoh, and T. Taniyama,  Appl. Phys. Lett. \textbf{99}, 102506 (2011).
%
\bibitem{Binek-05} P. Borisov, A. Hochstrat, X. Chen, W. Kleemann, and C Binek, Phys. Rev. Lett. \textbf{94}, 117203 (2005).
%
\bibitem{Laukhin-06} V. Laukhin, V. Skumryev, X. Mart\'{i}, D. Hrabovsky, F. S\'{a}nchez, M. V. Garc\'{i}a-Cuenca, C. Ferrater, M. Varela, U. L\"{u}ders, J.F. Bobo, and J. Fontcuberta, Phys. Rev. Lett. \textbf{97}, 227201 (2006).
%
\bibitem {Wu-10} S. M. Wu, Shane A. Cybart, P.  Yu, M. D. Rossell, J. X. Zhang, R. Ramesh, and R. C. Dynes, Nat. Material \textbf{9}, 756 (2010).

\bibitem{Berakdar-11} C.-L. Jia,  A. Sukhov,  P. P. Horley,  and J. Berakdar, J. Phys. Conf.
Ser. \textbf{303}, 012061 (2011).
%
\bibitem{Berakdar-12} P. P. Horley,  A. Sukhov,  C.-L. Jia,  E. Mart\'{i}nez,  and J. Berakdar, Phys. Rev. B \textbf{85}, 054401 (2012).
%
\bibitem{Bruchhausen-08} A. Bruchhausen,  A. Fainstein,  A. Soukiassian,  D. G. Schlom, X. X. Xi, M. Bernhagen,  P. Reiche,  and R. Uecker, Phys. Rev. Lett. \textbf{101}, 197404 (2008).
%
\bibitem{Piezoe1} G. Ederer and N. Spaldin, Phys. Rev. Lett. \textbf{95}, 257601 (2005).
%
\bibitem{Piezoe2} Richard M. Martin, Phys. Rev. B \textbf{5}, 1607 (1972).
%
\bibitem{Book-FM1} S. Chikazumi, \emph{Physics of Ferromagnetism} (Oxford University Press, New York 2002).
%
\bibitem{Parameters-FE} J. Hlinka and P. M\'{a}rton, Phys. Rev. B \textbf{74}, 104104 (2006).
%
\bibitem{Book-FM2} J. M. D. Coey, \emph{Magnetism and Magnetic Materials} (Cambridge University Press, Cambridge, 2010).
%
\bibitem{Book-FE} K. Rabe, Ch. H. Ahn, and J.-M. Triscone (eds.), \emph{Physics of Ferroelectrics: A Modern Perpective} (Springer, Berlin, 2007).
%
\bibitem{Book-MC} David P. Landau and K. Binder, \emph{A Guide to Monte-Carlo Simulations in Statistical Physics} (Cambridge University Press, New York 2009)
%
\bibitem{FE-MC1} B. G. Potter, Jr., B. A. Tuttle, and V. Tikare,  AIP Conf. Proc. \textbf{535}, 173 (2000).
%
\bibitem{FE-MC2} X. S. Gao, J.-M. Liu, X. Y. Chen and Z. G. Liu, J. Appl. Phys.  \textbf{88}, 4250 (2000).
%
\bibitem{Hinzke-98} D. Hinzke and U. Nowak, Phys. Rev. B \textbf{58}, 265 (1998).

%
\bibitem{SGL-93} P.  A. Serena, N. Garc\'{i}a, and A. Levanyuk, Phys. Rev. B \textbf{47}, 5027 (1993).
%
\bibitem{FM-MC1} M. Rao, H. R. Krishnamurthy, and R. Pandit, Phys. Rev. B \textbf{42}, 265 (856).
%
\bibitem{FM-MC2} M.  Acharyya and B. K. Chakrabarti, Phys. Rev. B \textbf{52}, 6500 (1995).
%
\bibitem{FM-MC3} X. Chen, Y. B. Guo, H. Yu, and J.-M. Liu, J. Appl. Phys. \textbf{100}, 103905 (2006).
%
\bibitem{v-BTO} Gerhard Mader, Hans Meixner, and Peter Kleinschmidt, J. Appl. Phys. \textbf{58}, 702 (1985).

\end{thebibliography}
\end{document}